\documentclass[
    ,final            % use final for the camera ready runs
  ]
  {aipproc}

\layoutstyle{8x11double}

\begin{document}

\title{Dynamics in a Bistable-Element-Network with Delayed Coupling and Local Noise}

\author{Daniel Huber and Lev Tsimring}{
  address={Institute for Nonlinear Science, University of California, San Diego, La Jolla, CA 92093-0402}
}

%\author{Lev Tsimring}{
%  address={Institute for Nonlinear Science, University of California, San Diego, La Jolla, CA 92093-0402}
%}

\begin{abstract}
The dynamics of an ensemble of bistable elements under the influence
of noise and with global time-delayed coupling is studied numerically
by using a Langevin description and analytically by using 1) a
Gaussian approximation and 2) a dichotomic model. We find that for a
strong enough positive feedback the system undergoes a phase
transition and adopts a non-zero stationary mean field. A variety of
coexisting oscillatory mean field states are found for positive and
negative couplings. The magnitude of the oscillatory states is maximal
for a certain noise temperature, i.e., the system demonstrates the
phenomenon of coherence resonance. While away form the transition
points the system dynamics is well described by the Gaussian
approximation, near the bifurcations it is more adequately described by
the dichotomic model.
\end{abstract}

\maketitle

%%%%%%%%%%%%%%%%%%%%%%%%%%%%%%%%%%%%%%%%%%%%
%% MAINMATTER
%%%%%%%%%%%%%%%%%%%%%%%%%%%%%%%%%%%%%%%%%%%%

\section{Introduction}

Stochastic rate processes in bi- or multi-stable systems lead to
many interesting phenomena observed in various scientific areas ranging
from physics to social science, and have thus been studied for a long
time. 

Here we consider a network of stochastically driven bistable elements
whose distinct feature is a time-delayed coupling. The time delays are
considered as uniform and the network elements are assumed to be
highly interconnected, so that the connectivity can be approximated by
a global all to all coupling.

The dynamics of the network is numerically explored by using a Langevin
model and analytically by using a Gaussian approximation and
a dichotomous model, derived from the corresponding Fokker-Planck equations
and Master equation, respectively.

\section{Langevin model}

The Langevin model consists of $N$ equations, each describing the
overdamped motion of a particle in a bistable potential in the
presence of noise and global coupling to a time-delayed mean field
$X(t-\tau)=N^{-1}\sum_{i=1}^Nx_i(t-\tau)$,
\begin{equation}
\dot{x}_i(t)= x_i(t)-x_i(t)^3+\varepsilon X(t-\tau)+\sqrt{2D}\xi(t),
\label{lang}
\end{equation}
where $\tau$ is the time delay, $\varepsilon$ is the coupling strength
of the feedback and $D$ denotes the variance of the Gaussian
fluctuations $\xi(t)$.

\section{Gaussian Approximation}

In order to theoretically study the dynamical properties of a globally
coupled set of noisy bistable elements (with no time delay), Desai and
Zwanzig \cite{Desai78} derived a hierarchy of equations for the
cumulant moments of the distribution function from the
multi-dimensional Fokker-Planck equation for the joint probability
distribution for all elements. For large noise intensities, when the
statistics of individual elements are approximately Gaussian, this
hierarchy can be truncated. Applying this approach to our system
yields the following set of equations for the mean field $X$ and the
variance $M=N^{-1}\sum(x_i-X)^2$,

\parbox{6.9cm}{
\begin{eqnarray*}
\dot{X}&=& X-X^3-3XM+\varepsilon X(t-\tau),
\label{XM}\\
\frac{1}{2}\dot{M}&=& M-3X^2M-3M^2+D.
\end{eqnarray*}
}\parbox{8mm}{\begin{eqnarray}\end{eqnarray}}

\section{Dichotomous Model}

To study the dynamics of a single bistable element with time-delayed
feedback Tsimring and Pikovsky \cite{Tsimring01} used a dichotomous
approximation. In the dichotomous approximation intra-well
fluctuations of $x_i$ are neglected, so that a bistable element can be
replaced by a discrete two-state system. Applying this complementary
approach to our system, allows us to express the mean field dynamics
in terms of the hopping rates $p_{12,21}$ which denote the probability
of a bistable element to change its state form $-1$ to $+1$ and vice
versa, respectively.  The equation for the mean field reads
\begin{equation}
\dot X = p_{12}-p_{21}-(p_{21}+p_{12})X,
\label{XX1}
\end{equation}
where the hopping probabilities are given by the Kramers transition
rate \citep{Kramers40}, which in the limit of small noise and small
coupling strength reads (cf. \cite{Tsimring01})
\begin{equation}
p_{12,21}=\frac{\sqrt{2\mp 3\varepsilon X(t-\tau)}}{2\pi}
\exp\left(-\frac{1\mp4\varepsilon X(t-\tau)}{4 D}\right). 
%p_{12,21}=\frac{\sqrt{2\mp 3\varepsilon X(t-\tau)}}{2\pi}
%{\rm e}^{-\frac{1\mp4\varepsilon X(t-\tau)}{4 D}}. 
\label{p1221}
\end{equation}

\section{Phase Diagram}

\begin{figure}
\includegraphics[width=5.3cm]{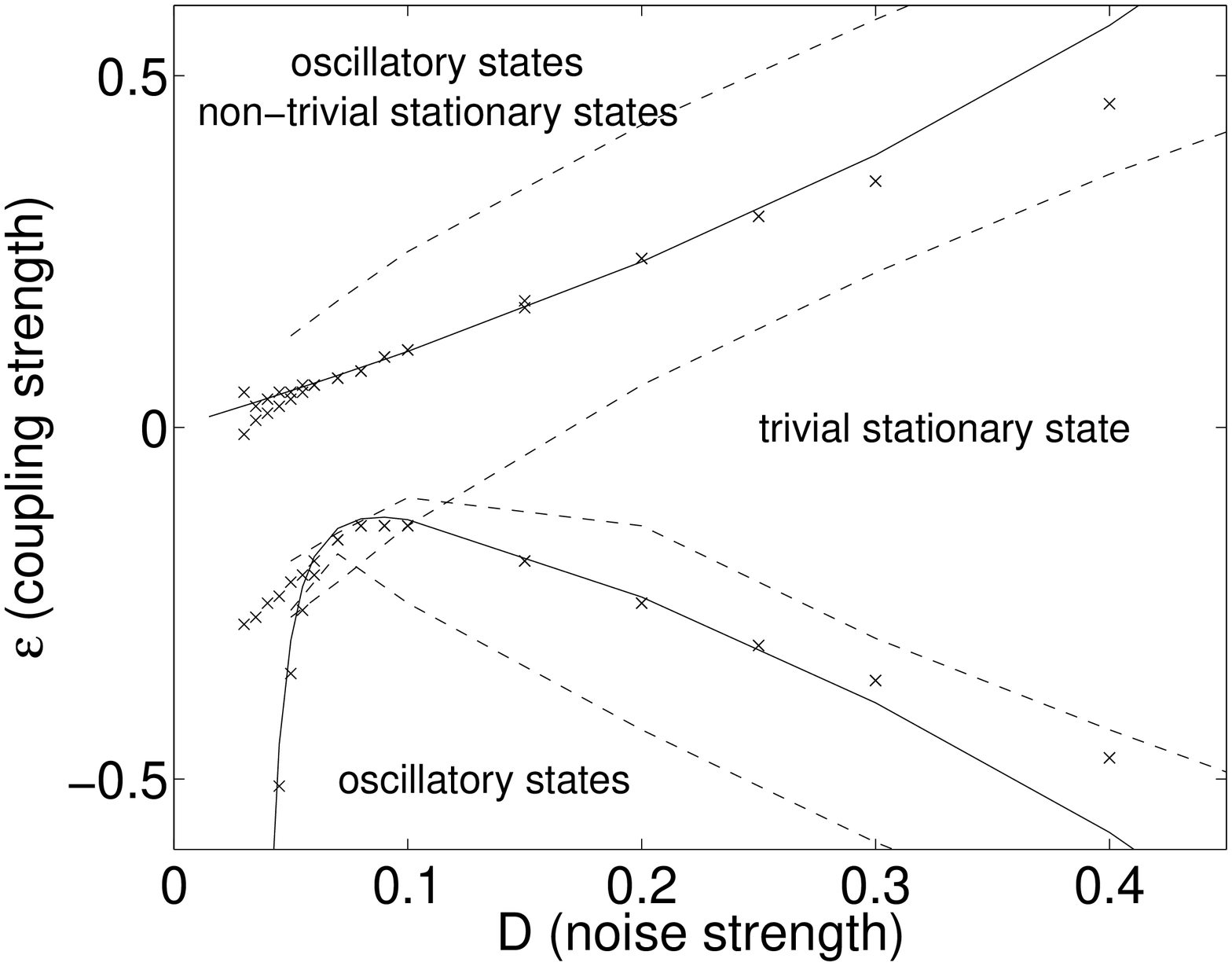}
\caption{\label{phase} { Phase diagram of the 
Langevin model (crosses), the Gaussian approximation (dashed lines) and
the dichotomous theory (solid lines). Double lines indicate hysteretic
transitions among phases.}}
\end{figure}

A numerical study of the Langevin model (Eq. \ref{lang}) shows that the system
undergoes ordering transitions and demonstrates multistability. That
is, for a strong enough positive coupling the system exhibits a
non-zero stationary mean field and a variety of stable oscillatory
states are accessible for positive and negative feedback. While the
transition to the non-zero stationary mean field is second order
(continuous), the type of the oscillatory transitions depends on the
system parameters and can be first order (discontinuous), associated
with hysteretic behavior, or second order. 

A linear stability analysis of Eq.~(\ref{XX1}) yields the critical
coupling strengths $\epsilon$ and the frequencies $\omega$ of the
accessible oscillatory states. It is found that the oscillation
periods can be arbitrary small meaning that a separation of time
scales of the macroscopic mean field dynamics and the microscopic
hopping dynamics of the constituent bistable elements is not possible.

A comparison of the Langevin model with the Gaussian approximation and
the dichotomous theory shows that while near the transition points the
system dynamics is strongly non-Gaussian, it is in this regime 
well described by a two-state model (see Fig.~\ref{phase}),
which allows for a complete analysis of the bifurcations of the
trivial equilibrium.

\section{Coherence Resonance}

\begin{figure}
\includegraphics[width=5.9cm]{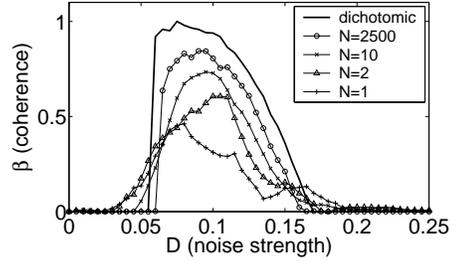}
\caption{\label{coherence} The normalized coherence $\beta$ of 
the oscillatory state at 
$\epsilon=-0.2.$ as a function of the noise strength D, for networks
of different size $N$ as well as for the dichotomic theory. The
coherence is given through $\beta=H\omega_p/\delta\omega$, where $H$
is the hight of the dominant spectral peak at $\omega_p$ and
$\delta\omega$ is the half-width of the peak.}
\end{figure}

The considered system exhibits the phenomena of coherence
resonance and array-enhanced resonance (see Fig. \ref{coherence}).

Both, Kramers random switching frequency $p$ (see \ref{p1221})
and the frequency of the oscillatory states $\omega$, resulting from
the coupling with the time-delayed mean field, depend on the noise
strength, i.e., $p=p(D)$ and $\omega=\omega(D)$. Thus, the
noise can be tuned so that the random hoping between the potential
wells of the bistable oscillators is synchronized with the periodic
modulation of the mean-field. This statistical synchronization takes
place when $\omega=\pi p$, where the coherence of the oscillatory states
becomes maximal.   

For the $N=1$ case the coherence resonance was observed by
\citet{Tsimring01}. We observe that the resonance phenomenon not only
persists in globally coupled networks with large $N$, but is enhanced, a
property which was found in other systems and is sometimes referred to
as array-enhanced resonance \citep{Zhou01}.

\begin{theacknowledgments}
This work was supported by the Swiss National Science Foundation
(D.H.) and by the U.S. Department of Energy, Office of Basic Energy
Sciences under grant DE-FG-03-96ER14592 (L.T.).
\end{theacknowledgments}

%\bibliography{mn-jour,dyn1}

\end{document}